\documentclass[twocolumn,
superscriptaddress,
nofootinbib,
amsmath,amssymb,
aps,
notitlepage
]{revtex4-1}

\usepackage{amsfonts,amsmath,units,wasysym,epsfig,graphicx,verbatim,color,subfigure,graphicx}

\usepackage{bibunits}

\usepackage[normalem]{ulem}
\usepackage{multirow}
\RequirePackage[colorlinks,citecolor=blue,urlcolor=magenta,linkcolor=blue]{hyperref}
\allowdisplaybreaks
\usepackage{orcidlink}
\begin{document}

\title{\bf Rotating black holes experience dynamical tides}

\author{Rajendra Prasad Bhatt\orcidlink{0009-0004-9088-2998}}
\email{rajendra@iucaa.in}
\affiliation{Inter-University Centre for Astronomy and Astrophysics, Pune 411007, India}
\author{Sumanta Chakraborty\orcidlink{0000-0003-3343-3227}}
\email{tpsc@iacs.res.in}
\affiliation{School of Physical Sciences, Indian Association for the Cultivation of Science, Kolkata-700032, India}
\author{Sukanta Bose\orcidlink{0000-0002-4151-1347}}
\email{sukanta@wsu.edu}
\affiliation{Department of Physics and Astronomy, Washington State University, 1245 Webster, Pullman, Washington 99164-2814, USA}


\begin{abstract} 
We find the {\em dynamical} tidal response of a Kerr black hole (BH) 
and demonstrate its tidal Love numbers to be non-vanishing when present in a non-axisymmetric external tidal field. To leading order, they depend quadratically on the black hole spin and linearly on the mode frequency.
This implies that Kerr BHs are deformable under certain external, time-dependent perturbations. 
Since non-vanishing tidal Love numbers have been used in compact 
binary coalescences 
observed by the LIGO-Virgo-KAGRA (LVK) Collaboration to infer their non-BH nature,
our findings have important implications on such inferences 
from future gravitational wave observations. 

\end{abstract}

\maketitle

{\bf Introduction.---}~Recent observations of 
gravitational waves (GWs) from 
binary black hole mergers~\cite{LIGOScientific:2016aoc, KAGRA:2021vkt} and 
images of 
shadows cast by 
supermassive black holes (BHs)~\cite{EventHorizonTelescope:2019dse, EventHorizonTelescope:2022wkp} have provided interesting tests of the strong gravity regime in an unprecedented manner.
They have also given a significant boost to the theoretical study of 
compact objects 
by ushering in new ways 
of probing their properties  with increasing precision. The response of a compact object to an external (tidal) field is one such property.
Here, we 
focus our attention on rotating BHs in general relativity, i.e., Kerr BHs. 
This Letter is an extension of our previous work~\cite{Bhatt:2023zsy}, where we 
studied the tidal response of Schwarzschild as well as slowly rotating Kerr BHs.

Tidal response of a self-gravitating body can be divided into two parts.
The first part is related to its deformation due to a tidal environment and is conservative. The second part is the effect of
tidal dissipation~\cite{Chia:2020yla}. One can associate dimensionless numbers with both 
parts, respectively termed as 
the tidal Love numbers (TLNs) and  dissipation numbers~\cite{Hinderer:2007mb, Flanagan:2007ix, Damour:2009vw, Binnington:2009bb, Chia:2020yla, Bhatt:2023zsy}.   
The TLNs of the
Schwarzschild black hole have been found to be zero in various studies~\cite{Binnington:2009bb, Damour:2009vw, Kol:2011vg, Chakrabarti:2013lua, Gurlebeck:2015xpa, LeTiec:2020spy, LeTiec:2020bos, Chia:2020yla, Charalambous:2021mea, Hui:2020xxx, Creci:2021rkz, Bhatt:2023zsy, Sharma:2024hlz}. Similar results have also been found for slowly rotating Kerr BHs~\cite{Landry:2015zfa, Pani:2015hfa, Chia:2020yla, Bhatt:2023zsy}. Non-linearities~\cite{DeLuca:2023mio} and stability~\cite{Katagiri:2023yzm} of the tidal response of non-rotating BHs have been studied too.
 
Contrastingly, Refs.~\cite{LeTiec:2020bos, LeTiec:2020spy} found non-zero and purely imaginary static TLNs for rotating BHs in non-axisymmetric tidal backgrounds.
They did so 
by using the Teukolsky equation
that describes the behavior of gravitational wave perturbations of the Kerr BH. The perturbation was modeled by 
an appropriate Newman-Penrose scalar. 
Reference~\cite{Chia:2020yla} 
extended those calculations and noted that the purely imaginary tidal Love numbers 
found in ~\cite{LeTiec:2020bos, LeTiec:2020spy} are not associated with tidal deformation but with tidal dissipation. 
Recently, we showed~\cite{Bhatt:2023zsy} that some terms are missing in the analysis of Ref.~\cite{Chia:2020yla} and calculated the tidal response function of Schwarzschild and slowly rotating Kerr BHs. In this Letter, we deduce the tidal response 
of an arbitrarily rotating BH and related quantities. 
In particular, we show that in a non-axisymmetric external tidal field it possesses non-vanishing dynamical TLNs.

With the advent of a new generation of GW detectors, it becomes imperative to study the properties of compact objects carefully since they may bear an influence on the GWs emitted by their binaries. While the study of the tidal deformability of neutron stars is useful in inferring their equation of state~\cite{LIGOScientific:2017vwq}, it also assumes importance for BHs in regard to tests of the presence of horizons. One can presumably use it for comparing the black holes of General Relativity with exotic compact objects~\cite{Cardoso:2017cfl, Chakraborty:2023zed} and black holes in alternative theories of gravity~\cite{Cardoso:2017cfl, DeLuca:2022tkm, Katagiri:2023umb}, which have been found to have non-zero TLNs in some cases~\cite{Cardoso:2017cfl, Chakraborty:2023zed,DeLuca:2022tkm}. Non-zero TLNs have also been reported for black holes in higher dimensions (e.g., braneworld BH~\cite{Chakravarti:2018vlt}, Myers-Perry BH~\cite{Charalambous:2023jgq, Rodriguez:2023xjd}, higher dimensional Schwarzschild BHs~\cite{Hui:2020xxx}), lower-dimensional BHs~\cite{DeLuca:2024ufn, Bhatt:2024mvr}, Schwarzschild-de Sitter BH~\cite{Nair:2024mya}, and for area quantized BHs~\cite{Nair:2022xfm}. Therefore, the study of TLNs can help us understand the various types of compact objects and might also be useful in studies of quantum effects in gravity. Below, we use $G=1=c$ and the positive signature metric. 

{\bf Executive summary.---}~The main results of this Letter are summarized below: 
(a) A Schwarzschild BH has vanishing static as well as dynamical TLNs; (b) A Kerr BH, both in the non-extremal and extremal limit, has vanishing static TLNs, but non-zero dynamical TLNs. The exact expressions for the non-zero dynamical TLNs have also been presented, for the first time. The dynamical TLNs vanish for axi-symmetric perturbations for both non-extremal and extremal BHs and also in the slow-rotation limit. Thus, our results address all possible TLNs for BHs in vacuum general relativity, within the linear-in-frequency approximation scheme.

Implications of this result for GW astrophysics and tests of gravity are potentially far-reaching. All binary BH waveform models assume zero TLNs for BHs, while estimating BH parameters and in testing deviations from the predictions of general relativity. 
Our results imply the presence of a systematic deviation in the waveform phase, which will make its presence known as more binary BHs are detected and the power of the above tests improves. If this deviation is not accounted for, it will result in incorrect deduction of astrophysical parameters, e.g., the mass and spin demographics of black holes, and impact the 
tests of general relativity.

{\bf Tidal signals in gravitational waves.---}~Originally the tidal response  was defined in  Newtonian gravity, e.g., when studying the multipolar deformation of a static, spherically symmetric, non-rotating body of mass $M$ and radius $\mathcal{R}$ subjected to a tidal environment. Here, the gravitational potential at a position $r$ from the center of mass of the object  
is the sum of (a) the undeformed object's potential, 
(b) the potential of the multipolar deformation, which is caused by the object's response to the tidal field, and (c) the potential of the ambient tidal field~\cite{LeTiec:2020bos, poisson_will_2014, Bhatt:2023zsy} (see Supplemental Material~\cite{supp_mat} for further details). 
In the frequency domain, this potential is
\begin{equation}
U=\frac{M}{r}-\sum_{lm}\frac{(l-2)!}{l!}\mathcal{E}_{lm}r^l\left[1+F_{lm}(\omega)\left(\frac{\mathcal{R}}{r}\right)^{2l+1}\right]Y_{lm}~,
\label{total_potential_Fourier_space}
\end{equation}
where 
the sum is over $l \in [2,\infty )$ and $m\in [-l,l]$,
$\omega$ is the angular mode frequency,
and
\begin{equation}
\label{response2tlns}
F_{lm}=2k_{lm}+i\omega\tau_0\nu_{lm}+\mathcal{O}(\omega^{2})
\end{equation}
is the tidal response function. 
$\mathcal{O}(\omega^{2})$ terms are neglected since we are working with the linear response and have taken the tidal field to be varying weakly with time. The above expression introduces three physical quantities --- (a) the TLNs $k_{lm}$, (b) the tidal dissipation numbers $\nu_{lm}$, and (c) the viscosity induced time-delay $\tau_0$~\cite{Chia:2020yla, Bhatt:2023zsy}.
This formalism can also be applied to slowly rotating self-gravitating bodies, with $\omega$ replaced by $\omega'$, which is the mode frequency in the body's co-rotating frame of reference~\cite{poisson_will_2014, Charalambous:2021mea}. 

So far the discussion was purely Newtonian and hence the results are
prone to gauge ambiguities when directly applied to relativistic systems~\cite{Gralla:2017djj}. In order to determine the TLNs and hence
the tidal response function in a gauge invariant manner, as argued in
Refs.~\cite{LeTiec:2020bos, Bhatt:2023zsy}, the Newman-Penrose scalar
$\Psi_{4}$ associated with the gravitational perturbation is an
appropriate quantity to analyze. In particular, $\Psi_{4}$ is directly
related to the Newtonian potential in the Newtonian
limit~\cite{LeTiec:2020bos, Bhatt:2023zsy}. Therefore, to deduce the tidal response
function in the relativistic context, one should determine $\Psi_4$ by
solving the Teukolsky equation and then 
examine its large $r$ behavior,
which will have contributions from positive as well as negative powers of $r$.
The coefficient of the $r^{-2l-1}$ term then determines the tidal response function $F_{lm}$, whose real part is the TLN associated with a given $(l,m)$ mode. While determining the TLNs, we will relax the domain of $l$ and allow it to be any complex number
in order to unambiguously identify the contributions from tidal deformation and dissipation~\cite{LeTiec:2020bos}. However, since in the end we are interested in the gravitational perturbation, the final answer will be gotten by 
restricting $l$ at the end of the calculation to obey $l\in\mathbb{Z}_{\ge2}$, which defines the set of positive integers greater
than or equal to 2. A variant of the above procedure will be used to calculate the tidal response function of an extremal Kerr black hole in an external tidal field. 

{\bf The near-zone Teukolsky equation.---}~We introduce the
re-scaled radial coordinate $z\equiv (r-r_+)/(r_+-r_-)$, which is
dimensionless and has the event horizon at its origin,\footnote{The extremal Kerr BH obeys $r_{+}=r_{-}$ and will be addressed separately below since $z$ diverges in this case.} 
and $R$ as the function that determines the radial dependence of $\Psi_4$. 
In terms of $z$, applying the near-zone condition ($M\omega z\ll 1$) and small frequency approximation $(M\omega\ll1)$, the Teukolsky equation for $R$
in the in-going Kerr coordinate simplifies to~\cite{Bhatt:2023zsy, Chakraborty:2023zed} (see Supplemental Material~\cite{supp_mat} for further details): 
\begin{equation}\label{GMATE}
\begin{split}
\frac{\mathrm{d}^2R}{\mathrm{d}z^2} 
&+ \left[\frac{2iP_+-1}{z} - \frac{2iP_1+1}{1+z}\right]\frac{\mathrm{d}R}{\mathrm{d}z}  \\
&+\Bigg[-\frac{4iP_+}{z^2}+\frac{4iP_2}{(z+1)^2}-\frac{l(l+1)-2}{z(1+z)}
\\
&+\frac{2am\omega}{z(1+z)}\left\{1+\frac{4}{l(l+1)}\right\} - \frac{2i\omega r_+}{z(1+z)}\Bigg]R=0\,,
\end{split}
\end{equation}
where $P_\pm = (am-2r_\pm M\omega)/(r_+-r_-)$, $P_1 = P_{+}+2\omega r_{+}$, and $P_2 = P_{+}+(\omega/2) (r_{+}+3r_{-})$.
We next solve the above master equation for an arbitrarily rotating black hole and obtain its tidal response function from the fall-off of the Weyl scalar in the intermediate region. 

{\bf Tidal response of a spinning black hole.---}~
Since Eq.~(\ref{GMATE}) is a second order differential equation and has three regular singular points at $z = 0,\,-1,\text{ and }\infty$, its solution can be written in terms of the Gauss hypergeometric function as
\begin{equation}
\begin{split}
R(z) &= c_1 z^{2} (z+1)^{2-N_3} \\ &\times\, _2F_1\left(3+l-N_2,2-l-N_1;3+2 i P_+;-z\right)\,,
\end{split}
\end{equation}
where $c_1$ is a constant of integration.\footnote{The second integration constant is fixed by invoking purely ingoing boundary condition at the event horizon.} 
The arguments of the hypergeometric function and the power of $(1+z)$ are written upto linear order in $M\omega$. For notational simplicity, the radial Teukolsky function is given in terms of three quantities, $N_1$, $N_2$, and $N_3$, all of which are linear functions of $\omega$ and are described in Supplemental Material~\cite{supp_mat}.\footnote{In the limit of non-rotating and slowly rotating BHs, they are
identical to $Q_1$, $Q_2$, and $Q_3$, respectively, defined in
Ref.~\cite{Bhatt:2023zsy}.}
The above result reduces to
the one for non-rotating and slowly rotating BHs~\cite{Bhatt:2023zsy}.

Given the radial Teukolsky function, which is purely ingoing at the BH
horizon, in order to determine tidal effects one computes the radial
part of $\Psi_{4}$ in the intermediate
regime. This is obtained by taking the large $r$ (or, equivalently, large
$z$) limit (see Supplemental Material~\cite{supp_mat} for further details).
We note the existence of the $z^{-2l-1}$ term in the large $r$ limit of radial $\Psi_4$, albeit with a frequency-dependent correction such that, upto linear order in $M\omega$, the fall-off behavior of the tidal response with  $z$ becomes $z^{-2l-1}\{1+(N_{2}-N_{1})\ln z\}$.
Thus, in the dynamical context, the leading behavior is as desired, with an additional $\ln z$ correction in the frequency. 
Following~\cite{Mandal:2023hqa, Perry:2023wmm} one can either retain the $\ln z$ term in the response function or 
define the response function without the $\ln z$ term~\cite{Katagiri:2023umb}.  
We choose the latter option\footnote{Note that the logarithmic term is often considered as an ambiguity term arising from differentiating the response of the body with the source of tidal perturbation. In this work, following~\cite{Katagiri:2023umb}, we have ignored the logarithmic term. Had we kept the term following the world-like effective field theory formalism, the results would be dependent on the cutoff scale. Connecting the TLNs derived here with GW observables will leverage such a scale dependence of the TLNs.} and obtain the tidal response function to be
\begin{align}\label{resp_func_arb_rot}
F_{lm} &= \frac{ \Gamma \left(-2 l-N_1+N_2-1\right)\Gamma \left(l+2 i
    P_{+}+N_1+1\right) }
{\Gamma
  \left(2-l-N_1\right) \Gamma \left(-l+2 i P_++N_2\right)}
  \nonumber
  \\
&\quad\quad \times \frac{\Gamma \left(3+l-N_2\right)}
{\Gamma  \left(2 l+N_1-N_2+1\right)}  \,.
\end{align}
The above expression differs from \cite{Chia:2020yla} by the presence
of terms involving $N_1$ and $N_2$, both of which are linear order
in $M\omega$. Therefore, if we take the limit of both $N_1$ and $N_2$
going to zero, it can be verified that the above response function
correctly reproduces the result presented in
Ref.~\cite{Chia:2020yla}. It should be emphasized that the terms
$N_{1}$ and $N_{2}$ must be included in the above expression for
consistency since they are of $\mathcal{O}(M\omega)$. Hence, it is
the form given in Eq.~(\ref{resp_func_arb_rot}) that correctly depicts the tidal response function of an arbitrarily rotating BH when one limits it to linear order terms in the frequency.

The static limit ($\omega=0$) of the tidal response function can be simplified to
\begin{align}\label{static_TLNs}
F^{\rm static}_{lm}&=-\left(\frac{iam}{r_{+}-r_{-}}\right)\frac{\left(l-2\right)!\left(l+2\right)!}{\left(2 l+1\right)!\left(2l\right)!}
\nonumber
\\
&\qquad \qquad \times \prod_{j=1}^{l}\left[j^2+\left(\frac{2am}{r_{+}-r_{-}}\right)^2\right]\,.
\end{align}
The above correctly reproduces the results of~\cite{LeTiec:2020spy, LeTiec:2020bos}. Moreover, the static tidal response function is purely imaginary and hence the static TLNs of an arbitrarily rotating BH identically vanish. This shows that BHs in vacuum general relativity have vanishing static TLNs.

Since our analysis is valid only upto linear order in $M\omega$, we
should expand the Gamma functions in powers of $M\omega$.
Using the prescription given in Eq.~(\ref{response2tlns}), we deduce the
TLNs for the general Kerr BH to be
\begin{equation}\label{dynamic_TLNs}
k_{lm} = (am)^2\,k^{(0)}_{lm} + am\omega\,k^{(1)}_{lm} + \mathcal{O}(M^{2}\omega^{2})~,
\end{equation}
where the explicit forms of $k^{(0)}_{lm}$ and $k^{(1)}_{lm}$ are given in Supplemental Material~\cite{supp_mat} and both of them are independent of frequency $\omega$. The first thing to notice from the above expression is that they are non-zero, signaling dynamical deformation for Kerr BHs under external tidal field. Moreover, from the above expression it is clear that the dynamical TLNs vanish for a Schwarzschild black hole ($a=0$), slowly rotating Kerr BH, and axi-symmetric tidal perturbation ($m=0$). These results are consistent with the earlier findings in the literature \cite{Chia:2020yla, Bhatt:2023zsy}. Even though it is tempting to take $\omega\to 0$ limit of the above equation and argue that Kerr BHs have non-zero static TLNs, that is not a correct statement. This is because the above expression was derived on the sole premise that $\omega \neq 0$, and for strictly static tidal fields the real part of the response function indeed vanishes, as we have already demonstrated above. Hence Eq.~(\ref{dynamic_TLNs}) is applicable only when $\omega \neq 0$, yielding non-zero dynamical TLNs. The singular behavior of the $\omega \to 0$, or the static limit, in connection with the TLNs, has also appeared earlier in various contexts, e.g., in the case of magnetic TLNs \cite{Pani:2018inf}, as well as for ultra-compact objects \cite{Chakraborty:2023zed}. To the best of our knowledge, the above result demonstrates for the first time
that rotating BHs have non-zero dynamical TLNs. Since all physical processes associated with binary BH coalescence are intrinsically dynamical in nature and all astrophysical BHs are expected to have some spin, the importance of the above result cannot be overstated. 

\begin{figure}
    \centering
    \includegraphics[scale=0.75]{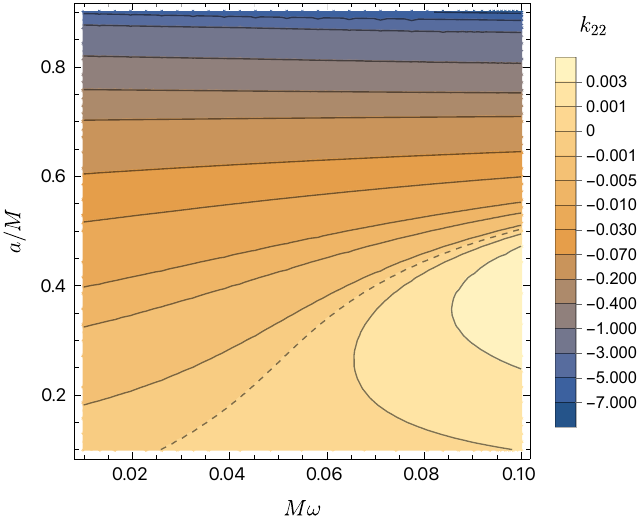}
    \caption{Contour plot of the tidal Love number $k_{22}$, as a function of $M\omega$ and $(a/M)$. As evident for larger values of the rotation parameter, the TLN becomes negative, while for smaller rotation and larger frequencies the TLN is positive. Intriguingly, for certain choices of frequency and rotation parameter, the TLN can be zero as well (shown by dashed contour).
    Note that our analysis is valid to linear order in $M\omega$; therefore, the $x-$axis has been plotted upto $M\omega =0.1$.}
    \label{fig_1}
\end{figure}

For a clearer depiction of the above result, we have presented the variation of the TLN associated with the $l=2=m$ mode ($k_{22}$) with the dimensionless mode frequency $M\omega$ and the dimensionless rotation parameter $(a/M)$ in Fig.~\ref{fig_1}. Evidently, for certain values of $M\omega$ and $(a/M)$, $k_{22}$ vanishes identically. Moreover, for larger values of $(a/M)$, $k_{22}$ is negative, while for smaller rotation and larger values of the frequency $M\omega$, $k_{22}$ becomes positive. Hence, rotation and frequency act oppositely in determining the sign of the TLN. Thus, for smaller mode frequency, $k_{22}$ of a rapidly rotating Kerr BH is negative, while for higher mode frequency, $k_{22}$ of a slowly rotating Kerr BH is positive. 

{\bf Tidal response of extremal Kerr black hole.---}~
While one might hope for using the results of the previous section to
also obtain the TLNs for an extremal Kerr BH -- by taking $a=M$ -- this simplistic idea does not work here:
The TLNs there explicitly depend on $(r_{+}-r_{-})^{-1}$, which
diverges in the extremal limit. This is simply due to
a bad choice of coordinates.
For this reason, we introduce the new coordinate $\bar{z}\equiv
(r-r_+)/r_+$, which merely shifts the origin of the radial coordinate to the position of the event horizon. 
Furthermore, since for an extremal Kerr BH $r_{+}=r_{-}=M$, one can express
$\Delta=(r-M)^{2}$, such that the source-free Teukolsky equation in
the new coordinate $\bar{z}$ is solvable in the near zone (see Supplemental Material~\cite{supp_mat} for further details). The
dynamical tidal
response function can be deduced from its solution as before, and is
found to be:
\begin{align}\label{non_static_rf_ext_BH}
F_{lm}&=-\{2i(m-2M\omega)\}^{2l+1-2N} \nonumber\\
& \quad\times \frac{\Gamma\left(-2l+2N\right) \Gamma\left(3+l-2iM\omega-N\right)}{\Gamma\left(2-l-2iM\omega+N\right)\Gamma\left(2l+2-2N\right)}
\end{align}
for an extremal Kerr BH under external tidal perturbation.
Here, $N \equiv 2mM\omega\left(l^2+l+4\right)/\{l(l+1)(2l+1)\}$ is a real quantity.
Note that, in the static limit ($\omega=0$), we obtain $N=0$ and, hence, the above dynamical tidal response function becomes
\begin{equation}
F_{lm}^{\rm static}=(-1)^{-1-l}im(2im)^{2l}\frac{\Gamma(3+l)\Gamma(-1+l)}{\Gamma(2l+1)\Gamma(2l+2)}\,,
\end{equation}
which is finite and purely imaginary, implying vanishing static TLNs for extremal Kerr BHs as well.

Given the fact that $N$ is real, one can read off the real part of the dynamical tidal response function in Eq.~(\ref{non_static_rf_ext_BH}) to obtain the dynamical TLNs,
\begin{equation}
k_{lm}=(2m)^{2l-2N-2}\left(m^2\Tilde{k}^{(0)}_{lm}+ mM\omega
\Tilde{k}^{(1)}_{lm}\right) + \mathcal{O}(M^2\omega^2)
\,,
\end{equation}
where  
the explicit forms of $\Tilde{k}^{(0)}_{lm}$ and $\Tilde{k}^{(1)}_{lm}$ are given in Supplemental Material~\cite{supp_mat} and show that 
they are non-zero for an extremal Kerr BH.

It
is also interesting to note that for an axi-symmetric tidal field
($m=0$), the above dynamical TLNs identically vanish. This can
be verified from the $m\to 0$ limit of the dynamical tidal response
function as well, on which we have elaborated in Supplemental 
Material~\cite{supp_mat}.
To
summarize, for extremal Kerr BHs, there are non-trivial TLNs,
which vanish in the static limit but yield
non-zero result in the dynamical scenario, akin to that of non-extremal Kerr BHs.

{\bf Discussion.---}~The tidal deformations of a compact object under an external gravitational field is a remarkable probe of its structure and composition.
They are encoded in the TLNs, which are imprinted on GWs radiated by it, e.g., when inspiraling in a binary. The measurement of the TLNs of the neutron stars in GW170817 constrained the equation of state of nuclear matter at extreme densities. It is now turning out that non-zero TLNs may not be exclusive to neutron stars; rather, the TLNs of ultra-compact objects depict a characteristic logarithmic behavior and BHs in higher dimensions have 
non-zero TLNs~\cite{Chakraborty:2023zed,Hui:2020xxx,Cardoso:2017cfl,Chakravarti:2018vlt}. However, BHs in General Relativity have vanishing static TLNs. It is an exciting prospect that at least some of these predictions can be tested in GW observations.

All of the above results were derived in the context of a static external tidal field, which does not depict the reality of an inspiraling binary. This motivates the pursuit of a dynamical description of tidal deformability, 
preferably in a gauge invariant way. Both of these objectives have been recently achieved through the Newman-Penrose scalar in the Teukolsky formalism~\cite{LeTiec:2020bos, Chia:2020yla}. The asymptotic expansion of the solution of the Teukolsky equation in the near-zone regime and under small-frequency approximation yields the dynamical tidal response function, whose real part yields the TLNs. As already pointed out in \cite{Bhatt:2023zsy} and also explicitly demonstrated here, the small-frequency approximation of \cite{Chia:2020yla} had missed quite a few terms, all of which are linear in $M\omega$. In this Letter we have kept all of these $\mathcal{O}(M\omega)$ terms and solved the Teukolsky equation to extract Gauss hypergeometric functions and, from their asymptotic limit, the dynamical tidal response function. Intriguingly, the static limit of this dynamical tidal response function yields a purely imaginary quantity, implying vanishing static TLNs for Kerr BHs. This holds true for extremal Kerr BHs as well. This leads to the first main result of this Letter, showing consistency of our formalism with previous literature:

\begin{enumerate}

\item  {\em Schwarzschild, slowly-rotating Kerr, non-extremal Kerr, and extremal Kerr BHs have vanishing \emph{static} TLNs. In other words, black holes in vacuum general relativity have vanishing \emph{static} TLNs.}
\end{enumerate}

\noindent Progressing to dynamical TLNs, we first proved that:
\begin{enumerate}
\setcounter{enumi}{1}
\item {\em Schwarzschild ($a=0$) and slowly-rotating Kerr BHs (i.e., with 
negligible $\mathcal{O}(a\omega)$ and $\mathcal{O}[(a/M)^2]$ terms) 
have vanishing \emph{dynamical} TLNs.}
\item {\em Both non-extremal and extremal Kerr BHs have vanishing \emph{dynamical} TLNs for axi-symmetric external tidal perturbations (i.e., with $m=0$).}
\end{enumerate}
\noindent In all cases in items 2 and 3, 
the dynamical tidal response function was shown to be purely imaginary.
 
Finally, in the case of arbitrarily rotating BHs, under non-axisymmetric tidal perturbation, the dynamical tidal response function is non-zero and has a real part, which has contributions from terms $\mathcal{O}(am\omega)$ as well as $\mathcal{O}(a^{2}m^{2})$. Given these parameter dependencies of the dynamical TLNs, it is obvious that the  limiting cases in items 2 and 3 are readily satisfied. 
A similar pattern of parameter dependencies holds for extremal Kerr BHs, with 
$a$ set to $M$.
Therefore, our final result is:

\begin{enumerate}
\setcounter{enumi}{3}
\item {\em Kerr BHs -- both non-extremal and extremal -- have non-vanishing \emph{dynamical} TLNs, within the linear-in-frequency approximation scheme,\footnote{However, we would like to emphasize that some other recent and ongoing works~\cite{Creci:2021rkz, Bhatt:2024rpx, Paper2} use different approximation schemes than the one presented here, and yet find results that are consistent with our findings, except for an overall normalization factor.} which depend on $(am)^{2}$ and $(am\omega)$, apart from other functions of the BH parameters.}

\end{enumerate}

Thus, the detection of a non-zero (dynamical) TLN via GWs from a binary with spinning components can not conclusively prove that the components are non-black holes, such as neutron stars.

We emphasize that our results are consistent with recent findings in the literature~\cite{Saketh:2023bul, Charalambous:2021mea}, for Ref.~\cite{Saketh:2023bul} showed that the linear-in-frequency TLNs for Kerr BHs have logarithmic running, suggesting non-zero dynamical TLNs, which we not only verified, but provided exact expressions for as well. The connection with GW waveforms can also be achieved through matched asymptotic expansion of $\Psi_{4}$ and then relating the flux at infinity with a GW phasing formula~\cite{Tichy:1999pv, Chatziioannou:2016kem}. In particular, it would be interesting to see if there are non-trivial effects on the parameter estimation from binary BH merger events, as the GW waveforms will be modified due to the non-zero dynamical TLNs, derived here.

There are several future directions of exploration. For example, the dependence of the non-zero dynamical TLNs on the spin of the perturbing field needs to be determined. In deriving the non-zero dynamical TLNs for a rotating BH, we employed perturbation theory. It would be interesting to see if the same result can be arrived at from other approaches, e.g., the ladder symmetry approach (we should be able to show that there is no ladder symmetry for gravitational perturbation of the Kerr BH in the dynamical context), or in effective field theory. Moreover, implications of the non-zero dynamical TLNs for extremal BH in the context of AdS/CFT correspondence will be worth exploring. We leave these issues for the future. 

{\bf Acknowledgements.---}~S.C.~acknowledges the warm hospitality at the Albert-Einstein Institute, where a part of this work was performed, which was supported by the Max-Planck-India mobility grant. Research of S.C. is supported by the MATRICS and the Core research grants from SERB, Government of India (Reg. Nos. MTR/2023/000049 and CRG/2023/000934). S.B. thanks IUCAA for its hospitality. Support from the National Science Foundation under Grant PHY-2309352 is acknowledged. 
The LIGO-Document number assigned to this document is LIGO-P2400258.

\bibliography{reference_arXiv}
\clearpage
\onecolumngrid
\setcounter{page}{1}
\setcounter{equation}{0}
\begin{bibunit}[apsrev4-1]

\def\eq#1{{Eq.~(\ref{#1})}}
\def\eqs#1{{Eqs.~(\ref{#1})}}
\def\sect#1{{Sec.~\ref{#1}}}
\def\EH{Einstein-Hilbert }
\def\LL{Lanczos-Lovelock }
\def\BY{Brown-York }
\def\gr{general relativity}
\def\RN{Reissner-Nordstr\"{o}m }
\def\KN{Kerr-Newmann }
\labelformat{section}{Section #1} 
\labelformat{subsection}{Section #1} 
\labelformat{subsubsection}{Section #1}
\labelformat{subsubsubsection}{Section #1}
\labelformat{equation}{Eq.~(#1)} 
\labelformat{figure}{Fig.~#1} 
\labelformat{subfigure}{Fig.~\thefigure#1} 
\labelformat{table}{Table~#1} 
\labelformat{appendix}{Appendix #1} 

\thispagestyle{empty}
\begin{centering}
{\large \bf 
Supplemental Material:\\
Rotating black holes experience dynamical tides}
\vspace{0.13 in}\\

Rajendra Prasad Bhatt\orcidlink{0009-0004-9088-2998},{\small $^{1}$} Sumanta Chakraborty\orcidlink{0000-0003-3343-3227},{\small $^{2}$} and Sukanta Bose\orcidlink{0000-0002-4151-1347}{\small $^{3}$}
\vspace{0.03 in}\\
{\small$^{1}$\em Inter-University Centre for Astronomy and Astrophysics, Pune 411007, India}\\
{\small$^{2}$\em School of Physical Sciences, Indian Association for the Cultivation of Science, Kolkata-700032, India}\\
{\small$^{3}$\em Department of Physics and Astronomy, Washington State University,\\ 1245 Webster, Pullman, Washington 99164-2814, USA}
\vspace{0.2 in}\\
\end{centering}

The purpose of this Supplemental Material is to describe the Teukolsky equation for the perturbation of a rotating black hole in an external tidal field and provide
 technical details of how it is solved to deduce a black hole's static and dynamical Love numbers, which are discussed in the main text. In the reference frame of an external observer, if the self-gravitating body and the tidal field are fixed, then it is called the static case. The tidal response function, and consequently the tidal Love numbers calculated for this case are known as static tidal response function and static tidal Love numbers, respectively. On the other hand, the self-gravitating body and the tidal field are moving with respect to each other in the dynamic case. Similarly, the tidal response function, and consequently the tidal Love numbers calculated for this case are known as dynamic tidal response function and dynamic tidal Love numbers, respectively. See \cite{Chakrabarti:2013lua_1} for an earlier work in this direction, however due to ambiguities the TLNs could not be determined uniquely.
We continue to
use $G=1=c$, 
unless otherwise stated. We also work with the 
positive signature metric; e.g., the Minkowski metric in Cartesian coordinate system will be given by $\text{diag.}(-1,+1,+1,+1)$. Moreover, $\mathbb{Z}_{\ge2}$ defines the set of positive integers greater than or equal to 2.

\section{Newtonian potential of a spherically symmetric object deformed in a tidal field}
Following the discussion of Eq.~(1) in the main text, the gravitational potential of a tidally deformed compact object is~\cite{LeTiec:2020bos_1, poisson_will_2014_1, Bhatt:2023zsy_1}
\begin{equation}
\label{total_potential}
U=U_{\text{body}}+U_{\text{tidal}}
=\frac{M}{r} 
+\sum_{l=2}^\infty \sum_{m=-l}^l \Big[\frac{(2l-1)!!}{l!}\frac{I_{lm}Y_{lm}}{r^{l+1}}
-\frac{(l-2)!}{l!}\mathcal{E}_{lm}Y_{lm}r^l\Big]\,,
\end{equation}
where the first term $(M/r)$ is the potential of the unperturbed spherically symmetric mass distribution while $I_{lm}$ are the multipole moments generated by the external gravitational field with tidal moments $\mathcal{E}_{lm}$~\cite{LeTiec:2020bos_1, poisson_will_2014_1, Bhatt:2023zsy_1}. Since $\mathcal{E}_{lm}$ are the cause for generating the multipole moments $I_{lm}$ in the originally static and spherically symmetric object, it is expected that they are related to each other~\cite{Chia:2020yla_1, Bhatt:2023zsy_1}:
\begin{equation}
I_{lm} = -\frac{(l-2)!}{(2l-1)!!}\left[2k_{lm}\mathcal{E}_{lm}-\tau_0\nu_{lm}\dot{\mathcal{E}}_{lm}+\cdots\right] \mathcal{R}^{2l+1}~.
\label{I_def}
\end{equation}
Note that $\mathcal{O}(\mathcal{E}_{lm}^{2})$ and  
$\mathcal{O}(\ddot{\mathcal{E}}_{lm})$ terms have been dropped since we are working with the linear response and have assumed the tidal field to be weakly varying with time. 
It is to be emphasized that the factor $\{(l-2)!/(2l-1)!!\}$ in front of the above expression on the right-hand side (RHS) is a normalization factor, and can differ from one work to another.
In the frequency domain, the above potential takes the expression given in Eq.~(1) of the main text. 

To determine the tidal response
function in the relativistic context, one obtains the Newman-Penrose scalar $\Psi_4$ by
solving the Teukolsky equation -- of the perturbed geometry of the desired black hole -- and then 
deduces its large $r$ behavior. Here, $r$ being large means that the field point is sufficiently far away from the deformed object yet sufficiently deep within the tidal field. This is referred to as the intermediate region. Throughout this letter, by large $r$ expansion we will imply extension to the intermediate regime. The asymptotic behavior, unless it is already present in the metric, never affects the TLN (see \cite{Nair:2024mya_1}). 
As a consequence $\Psi_4$ in the intermediate region will have contributions from the positive as well as negative powers of $r$:
\begin{equation}
\label{psi_4_intermediate}
\Psi_{4}^{\text{intermediate}}=\frac{1}{4}\sum_{l=2}^\infty \sum_{m=-l}^l
\sqrt{\frac{(l+2)(l+1)}{l(l-1)}}\,\mathcal{E}_{lm}\,r^{l-2}
\left[1+F_{lm}\left(\frac{\mathcal{R}}{r}\right)^{2l+1}\right]\,_{-2}Y_{lm}\,.
\end{equation}
Here, the coefficient of the $r^{-2l-1}$ term determines the tidal response function $F_{lm}$ (further details are given in the main text).

\section{Teukolsky equation in in-going Kerr coordinates}\label{sec:Master_Teukolsky_eq}
We start with the Teukolsky equation, governing the gravitational perturbations of the Kerr background~\cite{Teukolsky:1972my_1, Teukolsky:1973ha_1, Press:1973zz_1, Teukolsky:1974yv_1}, before discussing specific limits of this equation that we are interested in. For writing down the Teukolsky equation, we will use the in-going Kerr coordinates, i.e., $\{v,r,\theta,\widetilde{\phi}\}$.
The Kerr metric in the in-going Kerr coordinates~\cite{Chia:2020yla_1, Bhatt:2023zsy_1} and the relation between in-going Kerr coordinates and Boyer-Lindquist coordinates $\{t,r,\theta,\phi\}$ can be found in~\cite{Teukolsky:1974yv_1,  Bhatt:2023zsy_1} and hence we will not reproduce it here. Before writing down the Teukolsky equation for gravitational perturbation, we start by decomposing the Newman-Penrose scalar $\Psi_{4}$ into radial and angular parts as~\cite{Teukolsky:1974yv_1}
\begin{equation}\label{gen_eq_decomp}
\rho^{4}\Psi_{4}=\int \mathrm{d}\omega\,e^{-i\omega v}\sum_{lm} e^{-im\widetilde{\phi}}\,_{-2}S_{lm}(\theta)R(r)~,
\end{equation}
where $\rho\equiv-(r-ia\cos\theta)$. The angular part, $\,_{-2}S_{lm}(\theta)$, satisfies the equation for spin-weighted spheroidal harmonics~\cite{Teukolsky:1974yv_1}, while the radial function satisfies the radial Teukolsky equation, which in the in-going Kerr coordinate is given by~\cite{Teukolsky:1974yv_1, Chia:2020yla_1}
\begin{multline}\label{Teqr}
\dfrac{\mathrm{d}^2R}{\mathrm{d}r^2}+\left(\frac{2iP_+-1}{r-r_+}-\frac{2iP_-+1}{r-r_-}-2i\omega\right)\dfrac{\mathrm{d}R}{\mathrm{d}r} 
\\ 
+\left[-\frac{4iP_+}{(r-r_+)^2}+\frac{4iP_-}{(r-r_-)^2}+\frac{A_-}{(r-r_-)(r_+-r_-)}-\frac{A_+}{(r-r_+)(r_+-r_-)}\right]R=\frac{T}{\Delta}\,,
\end{multline}
where $P_+$ is defined in the main text, and $A_\pm = 2i\omega r_\pm +\lambda$.
In addition, $T$ is an appropriate source term and $\lambda\equiv E_{lm} -2am\omega +a^2\omega^2 -2$. Here, $E_{lm}$ refers to the eigenvalue of the angular equation and can be expanded in frequency as~\cite{Press:1973zz_1, Fackerell:1977_1, Seidel:1988ue_1, Berti:2005gp_1},
\begin{equation}
E_{lm}=l(l+1)-\left(\frac{8m}{l(l+1)}\right)a\omega+\mathcal{O}\left(a^{2}\omega^{2}\right)\,.
\end{equation}
Now we can introduce a coordinate transformation $z\equiv (r-r_+)/(r_+-r_-)$, and apply the near-zone condition ($M\omega z\ll 1$) and small frequency approximation $(M\omega\ll1)$, which transforms \ref{Teqr} to Eq. (3) of the main text.


\section{Tidal response of a non-extremal Kerr black hole}\label{supp_mat:sec_1}

Having derived the near-zone radial Teukolsky equation in the small frequency regime, in Eq.~(3) of the main text, here we present its solution and, from its asymptotic behavior, deduce the tidal response function of an arbitrarily rotating BH. 
The solution of Eq.~(3) of the main text is
\begin{multline}\label{radialfngen}
R(z)=(z+1)^{2-N_3} \left[c_1 z^{2} \, _2F_1\left(3+l-N_2,2-l-N_1;3+2 i P_+;-z\right)\right.\\\left.+c_2 z^{-2 i P_+} \, _2F_1\left(l-2 i P_+-N_2+1,-l-2 i P_+-N_1;-1-2 i P_+;-z\right)\right]\,,
\end{multline}
where $c_1$ and $c_2$ are the constants of integration. The arguments of the hypergeometric functions and the power of $(1+z)$ are written upto linear order in $M\omega$. For notational simplicity, we have expressed the solution of the radial Teukolsky function in terms of three quantities, $N_1$, $N_2$, and $N_3$, all of which are linear functions of the frequency $\omega$, and have the following expressions:
\begin{equation}\label{def_N1}
N_1 = 2 \omega \left[-\frac{4 a m}{l(l+1)(2l+1)}+\frac{- a m+4 i l M-i(2l-1)r_+}{2 l+1}+\frac{2 \left(r_+-M\right) \left(r_+-M+ia m\right)}{a m+2i (r_+-M)}+\frac{4 i \left(r_+-M\right)}{2 l+1}\right]\,,
\end{equation}
\begin{equation}\label{def_N2}
N_2= 2 \omega \left[\frac{4 a m}{l(l+1)(2l+1)}+\frac{a m+4 i (l+1) M-i (2 l+3) r_+}{2 l+1}+\frac{2 \left(r_+-M\right) \left(r_+-M+i a m\right)}{a m+2i (r_+-M)}-\frac{4 i \left(r_+-M\right)}{2 l+1}\right]\,,
\end{equation}
and
\begin{equation}
N_{3}=\frac{12 \omega \left(r_+-M\right){}^2}{a m+2i(r_{+}-M)}~.
\end{equation}
To proceed further, we first fix one of the arbitrary constants, among $c_{1}$ and $c_{2}$, by considering the near-horizon limit of \ref{radialfngen}. This is obtained by taking the $z\to 0$ limit.
Since there are no outgoing modes at the event horizon of a BH, we need to impose the condition $c_{2}=0$ in \ref{radialfngen} for the radial Teukolsky function, which then reduces to Eq.~(4) of the main text.
Now one can compute the radial part of the Newman-Penrose scalar $\Psi_{4}$ in the intermediate regime, which is obtained by taking large $r$ (or, equivalently large $z$) limit, and is given by,
\begin{multline}\label{psi4gen}
\Psi^{\rm intermediate}_{4}\propto z^{l-2+N_1 -N_3} \left\{\frac{\Gamma \left(3+2 i P_+\right) \Gamma \left(2 l+N_1-N_2+1\right)}{\Gamma \left(l+2 i P_++N_1+1\right) \Gamma \left(3+l-N_2\right)}\right\}\\\times\left[1+z^{-2l-1+N_2-N_1} \left\{\frac{ \Gamma \left(-2 l-N_1+N_2-1\right)\Gamma \left(l+2 i P_++N_1+1\right) \Gamma \left(3+l-N_2\right)}{\Gamma \left(2-l-N_1\right) \Gamma \left(-l+2 i P_++N_2\right)\Gamma \left(2 l+N_1-N_2+1\right) }\right\}\right]\,.
\end{multline}
In order to determine the tidal response function, one must compare the above expression
with the one presented in \ref{psi_4_intermediate}.
In this way, we obtain the tidal response function to be Eq.~(5) of the main text.
Our next objective is to simplify it.

The first step in the simplification process
follows from the reflection formula for  Gamma functions, i.e., $\Gamma(z)\Gamma(1-z)=\{\pi/\sin(\pi z)\}$. By employing it we obtain
\begin{equation}
\frac{\Gamma \left(-2 l-N_1+N_2-1\right)}{\Gamma \left(2-l-N_1\right)}=\alpha \left(\frac{\Gamma \left(-1+l+N_1\right)}{\Gamma \left(2 l+N_1-N_2+2\right)}\right),\qquad \frac{1}{\Gamma \left(-l+2 i P_++N_2\right)} = \xi\,\Gamma \left(1+l-2 i P_+-N_2\right)\,,
\end{equation}
where we introduced the two quantities 
\begin{equation}\label{def_alpha_s_xi_s}
\alpha\equiv\frac{\sin\pi(l-2+ N_1)}{\sin\pi(1 + 2 l + N_1 - N_2)}\,, 
\qquad \xi\equiv \frac{\sin\left[\pi(-l+2 i P_++N_2)\right]}{\pi}\,.
\end{equation}
Using the above Gamma function identities, we can re-express 
the tidal response function (Eq.~(5) of the main text) as
\begin{equation}\label{resp_func_arb_rot_1_s}
F_{lm} = \alpha\xi\frac{ \Gamma \left(-1+l+N_1\right)\Gamma \left(l+2 i P_++N_1+1\right) \Gamma \left(3+l-N_2\right)\Gamma \left(1+l-2 i P_+-N_2\right)}{\Gamma \left(2 l+N_1-N_2+2\right) \Gamma \left(2 l+N_1-N_2+1\right)}\,.
\end{equation}
Since $M\omega$ is small, we can expand each and every Gamma functions in the numerator and the denominator of \ref{resp_func_arb_rot_1_s} upto linear orders of $M\omega$. For this purpose, we may use the following Taylor series expansion of the Gamma functions~\cite{abramowitz_stegun_1972_1}:
\begin{equation}
\Gamma(f(z))=\Gamma(f(z_0)) + (z-z_0)\psi(f(z_0))\Gamma(f(z_0))\frac{\mathrm{d}f(z)}{\mathrm{d}z}\bigg|_{z=z_0} + \mathcal{O}[(z-z_0)^2]\,.
\end{equation}
Here $\psi(z)$ is the di-gamma function defined as $\psi(z) = \Gamma'(z)/\Gamma(z)$, where $\Gamma'(z)$ is the first order derivative of $\Gamma(z)$ with respect to the argument $z$. Using these results, the response function can be written as,
\begin{multline}\label{resp_func_arb_rot_1_expansion}
F_{lm}=\alpha\xi\left\{\frac{\Gamma \left(-1+l\right)\Gamma \left(l+2 i P_++1\right) \Gamma \left(3+l\right)\Gamma\left(1+l-2 i P_+\right)}{\Gamma \left(2 l+2\right)\Gamma \left(2 l+1\right)}\right\}
\Big[1+N_1\psi(-1 +l)+N_1\psi(1 + l+2 i P_+)\\- N_2\psi(3 + l)-N_2\psi(1 + l-2 i P_+)- (N_1 - N_2)\psi(2 + 2 l)- (N_1 - N_2)\psi(1 + 2 l)\Big]\,,
\end{multline}
where we have ignored the second and higher order terms of $M\omega$. To further simplify the above equation, we can use the following identity
(assuming $l$ to be an integer),
\begin{equation}
\Gamma(l+1+2iP_+)\Gamma(l+1-2iP_+)=\frac{2i\pi P_+}{\sin(2i\pi P_+)} \prod_{j=1}^{l} (j^2+4P_+^2)\,.
\end{equation}
Consequently, the response function becomes
\begin{align}\label{response_1}
F_{lm}&=(2i P_+)\alpha\,\kappa\,\frac{\Gamma \left(-1+l\right)\Gamma \left(3+l\right)}{\Gamma \left(2 l+2\right) \Gamma \left(2 l+1\right) }\prod_{j=1}^{l} (j^2+4P_+^2)
\Big[1+N_1\psi(-1 +l)+N_1\psi(1 + l+2 i P_+)
\nonumber
\\
&\qquad-N_2\psi(3 + l)-N_2\psi(1 + l-2 i P_+)-(N_1 - N_2)\psi(2 + 2 l)- (N_1 - N_2)\psi(1 + 2 l)\Big]\,,
\end{align}
where $\alpha$ is defined in \ref{def_alpha_s_xi_s} and 
\begin{equation}\label{def_kappa}
\kappa\equiv\frac{\sin(-l+2 i P_++N_2)\pi}{\sin(2i\pi P_+)}\,. 
\end{equation}

This is where the static and dynamical TLNs branch out. The static TLNs are determined by taking $\omega \to 0$ limit of the above response function and then obtaining its real part. In this case, $P_{+}=am/(r_{+}-r_{-})$, and the two quantities $\alpha$ and $\kappa$ reduce to
\begin{equation}
\alpha_0\equiv\lim_{\omega \rightarrow0}\alpha=\frac{(-1)^{l+1}}{2}\,,
\qquad
\kappa_{0}\equiv \lim_{\omega \rightarrow0}\kappa=(-1)^{l}-\sin(\pi l)\coth(2i\pi P_{+})\,,
\end{equation}
respectively.
Given the above values for $\alpha_{0}$ and $\kappa_{0}$, besides the other factors in \ref{response_1}, it follows that the response function is finite if one takes
$l$ to be an integer. Therefore, the static tidal response function becomes Eq.~(6) of the main text, which showed that it
is purely imaginary. Hence, the static TLNs of an arbitrarily rotating BH identically 
vanish.

On the other hand, for non-zero frequency, upto linear order in 
$M\omega$, and for
$l\in \mathbb{Z}_{\ge2}$, the quantities $\alpha$ and $\kappa$ simplify to yield the following results:
\begin{equation}
\alpha=(-1)^{l+1}\left(\frac{N_1}{N_1-N_2}\right)\,, 
\qquad 
\kappa=(-1)^l\left[1-i\pi N_2\coth(2 P_+\pi)\right]\,.
\end{equation}
Therefore, the dynamical tidal response function of an arbitrarily rotating BH takes the following form 
\begin{equation}\label{response_New1a}
F_{lm}=(-2iP_{+})\left(\frac{N_1}{N_1-N_2}\right)\Big[1-i\pi N_2\coth(2 P_+\pi)\Big]F_2\left(1+M\omega F_1\right)\,,
\end{equation}
where the ratio $\{N_{1}/(N_{1}-N_{2})\}$ is independent of $\omega$, and the quantities $F_{1}$ and $F_{2}$ are given by
\begin{equation}\label{def_Mo_F1}
M\omega F_{1}\equiv N_1\psi(-1 +l)+N_1\psi(1 + l+2 i P_+)- N_2\psi(3 + l)-N_2\psi(1 + l-2 i P_+)
-(N_1 - N_2)\psi(2 + 2 l)- (N_1 - N_2)\psi(1 + 2 l)
\end{equation}
and
\begin{equation}\label{def_F2}
F_{2}\equiv \frac{\left(l-2\right)!\left(l+2\right)!}{\left(2 l+1\right)!\left(2l\right)!}\prod_{j=1}^{l} (j^2+4P_+^2)\,. 
\end{equation}
Given the fact that $N_{1}$ and $N_{2}$ are themselves complex, it is not obvious that the above response function is purely complex. In what follows, we will explore the above tidal response function in more detail and discuss its various limits, before commenting on the vanishing/non-vanishing of dynamical TLNs for an arbitrarily rotating BH. Another point 
that must be emphasized 
is that the above response function is strictly valid for non-extremal Kerr BH. This is because in the extremal limit $P_{+}$ diverges. This is due to the choice of the coordinate $z$, which is ill-behaved in the $a\to M$ limit. This is why we have discussed the tidal response of an extremal Kerr BH separately.

\subsection{Dynamical tidal deformation and dissipation}
\label{sec:Tidal_response_arb_rot_BH_cal_real_Imag}

We have demonstrated above that in the static limit 
the TLNs
of an arbitrarily rotating (non-extremal) BH identically 
vanish.
In this section we will focus exclusively on BHs experiencing dynamical tides and examine 
their non-rotating as well as slowly-rotating limits, before studying the general case of an arbitrarily rotating BH. There are three specific cases to consider --- (a) Schwarzschild BH, obtained in the $a\to0$ limit of \ref{response_New1a}; (b) Slowly rotating BH, for which $a\ll M$, and finally (c) an arbitrarily rotating black hole in an axi-symmetric tidal environment, i.e., the $m\to0$ limit of \ref{response_New1a}. In all of these cases, $P_+$ 
is a small quantity. Since $N_1$ and $N_2$ are also of the $\mathcal{O}(M\omega)$, we can ignore 
terms like $P_+\,N_1$ and $P_+\,N_2$ in the tidal response function. Therefore, the tidal response function
presented in \ref{response_New1a}
simplifies drastically to
\begin{equation}
F^{\text{Sch, Kerr(slow, axi)}}_{lm}=(-2i P_+)\left(\frac{N_1}{N_1-N_2}\right)\Big[1-i\pi N_2\coth(2 P_+\pi)\Big]\frac{\left(l-2\right)!\left(l+2\right)!}{\left(2 l+1\right)!\left(2l\right)!}\prod_{j=1}^{l} (j^2)\,.
\end{equation}
Above, the factor $\coth (2\pi P_{+})$ can be 
written as
\begin{equation}\label{cothx}
\coth (2\pi P_+) = \frac{1}{2\pi P_+} + \frac{1}{\pi}\sum_{k=1}^{\infty}\left(\frac{4P_+}{k^2 +4(P_+)^2}\right)\,.
\end{equation}
Furthermore, since $P_{+}$ is small, we have expressed $\prod_{j=1}^{l}(j^2+4P_+^2) \sim \prod_{j=1}^{l}j^2=(l!)^2$, which yields the following expression for the tidal response function:
\begin{equation}
F^{\text{Sch, Kerr(slow, axi)}}_{lm}=-\left(\frac{N_1}{N_1 - N_2}\right)\left(2iP_{+}+N_2\right)\frac{\left(l-2\right)!\left(l+2\right)!(l!)^{2}}{\left(2 l+1\right)!\left(2l\right)!}\,.
\end{equation}
For Schwarzschild and slowly rotating Kerr BHs, one can verify that both $N_1$ and $N_2$ are purely imaginary and, hence, it follows that the tidal response function is also 
a
purely imaginary quantity. Therefore, 
the dynamical TLNs of Schwarzschild and slowly rotating Kerr BHs identically 
vanish,
which is consistent with the findings of Ref.~\cite{Bhatt:2023zsy_1}. 
Interestingly, even
for an arbitrarily rotating black hole in an axi-symmetric tidal background  
$N_1$ and $N_2$ are found to be purely imaginary, by substituting $m=0$ in the respective expressions 
in \ref{def_N1} and \ref{def_N2}.
Therefore, even in this case the dynamical TLNs identically vanish. All of  
this suggests 
that the dynamical TLNs of an arbitrarily rotating BH, if non-zero, must be proportional to $am\omega$, and this is precisely what we have demonstrated here, 
albeit, with
additional contribution coming from $(am)^2$.

At this stage, we note that while the response function given by \ref{response_1} matches the response function presented in~\cite{Bhatt:2023zsy_1} (in their respective limits), its simplified form presented in \ref{response_New1a} does not match it. 
This is  
because in Ref.~\cite{Bhatt:2023zsy_1}, we had assumed $l$ to be an integer at the very end of the computation, 
and not at an intermediate step as done here.
Despite this difference, the broad result remains identical, namely, that the dynamical TLNs of non-rotating and slowly rotating BHs identically vanish.  

Having dealt with all the limiting cases of interest, let us concentrate on the tidal response function of an arbitrarily rotating BH, whose real part provides information about the TLNs, while the imaginary part is related to tidal dissipation. In order to analyze the tidal response function of an arbitrarily rotating black hole, presented in \ref{response_New1a}, 
note that $F_{1}$ is a complex quantity while $F_{2}$ is a purely real quantity. Therefore, the real part of $F_{lm}$ will be determined by certain combinations of $N_{1}$, $N_{2}$ and $F_{1}$, and leads to 
\begin{equation}
k_{lm} \equiv \frac{1}{2}\textrm{Re} \left(F_{lm}\right)=-P_{+}F_{2}\,\text{Im}\Big[\left(\frac{N_1}{N_1-N_2}\right)\Big\{1-i\pi N_2\coth(2 P_+\pi)\Big\}\left(1+M\omega F_1\right)\Big]\,,
\end{equation}
where, for any complex quantity $\mathbb{C}$, 
we resolve it into its real part $\text{Re}(\mathbb{C})$ and imaginary part
$\text{Im}(\mathbb{C})$, as $\mathbb{C}=\text{Re}(\mathbb{C})+i\,\text{Im}(\mathbb{C})$. Thus, we can write the dynamical TLNs as
\begin{align}\label{eq_39_a}
k_{lm}&=-P_{+}F_{2}\Bigg[-\pi\coth(2 P_+\pi)\text{Re}\left(\frac{N_1}{N_1-N_2}\right)\text{Re}N_{2}
+\text{Im}\left(\frac{N_1}{N_1-N_2}\right)\left\{1+\pi \coth(2 P_+\pi)\textrm{Im}N_{2}\right\}
\nonumber
\\
&+\left\{\text{Re}\left(\frac{N_1}{N_1-N_2}\right)\left\{1+\pi \coth(2 P_+\pi)\textrm{Im}N_{2}\right\}+\pi\coth(2 P_+\pi)\text{Im}\left(\frac{N_1}{N_1-N_2}\right)\text{Re}N_{2}\right\}\text{Im}(M\omega F_1)
\nonumber
\\
&+\left\{-\pi\coth(2 P_+\pi)\text{Re}\left(\frac{N_1}{N_1-N_2}\right)\text{Re}N_{2}+\text{Im}\left(\frac{N_1}{N_1-N_2}\right)\left\{1+\pi \coth(2 P_+\pi)\textrm{Im}N_{2}\right\}\right\}\text{Re}(M\omega F_1)\Bigg]~.
\end{align}
To further simplify the expression for the dynamical TLNs of a non-extremal Kerr BH, 
first we will analyze the form of $N_1$ and $N_2$, as presented in \ref{def_N1} and \ref{def_N2}, respectively. The real and imaginary parts of $N_1$ and $N_2$ can be written as
\begin{equation}\label{ReImN1N2}
\text{Re}(N_1) = am\omega\,x_{N_1}, \qquad \text{Im}(N_1) = M\omega\,y_{N_1}, \qquad \text{Re}(N_2) = am\omega\,x_{N_2}, \qquad \text{Im}(N_2) = M\omega\,y_{N_2}\,,
\end{equation}
where the quantities $x_{N_1}$, $x_{N_2}$, $y_{N_1}$, and $y_{N_2}$ 
take
the following 
forms:
\begin{align}
x_{N_1}&=-\frac{2(l^2+l+4)}{l(l+1)(2l+1)}+\frac{12 (r_+-M)^2}{a^2m^2+4(r_+-M)^2}\,,
\label{ReN1_a}
\\
y_{N_1}&=\frac{2}{M} \left[\frac{4(l-1) M-(2 l-5) r_+}{2 l+1}-\frac{2 \left(r_+-M\right) \left\{-a^2m^2+2(r_+-M)^2\right\}}{a^2m^2+4(r_+-M)^2}\right]\,,
\label{ImN1_a}
\\
x_{N_2}&=\frac{2(l^2+l+4)}{l(l+1)(2l+1)}+\frac{12 (r_+-M)^2}{a^2m^2+4(r_+-M)^2}\,,
\label{ReN2_a}
\\
y_{N_2}&=  \frac{2}{M} \left[\frac{4 (l+2) M- (2 l+7) r_+}{2 l+1}-\frac{2 \left(r_+-M\right) \left\{-a^2m^2+2(r_+-M)^2\right\}}{a^2m^2+4(r_+-M)^2}\right]\,.
\label{ImN2_a}
\end{align}
Similarly, we can write the real and imaginary parts of the difference $(N_1-N_2)$ as
\begin{equation}\label{ReImN1mN2}
\text{Re}(N_1-N_2) = am\omega\,x_{N_{12}}\,, \qquad \text{Im}(N_1-N_2) = M\omega\,y_{N_{12}}\,,
\end{equation}
where
\begin{equation}\label{ReImN1mN2_a}
x_{N_{12}}=-\frac{4(l^2+l+4)}{l(l+1)(2l+1)}\,, \qquad y_{N_{12}}=\frac{24\left(r_+-M\right)}{(2 l+1)M}\,.
\end{equation}
Thus, it is clear that the real parts of $N_1$, $N_2$, and $(N_1-N_2)$ have $am\omega$ as a multiplication factor while the imaginary parts of $N_1$, $N_2$, and $(N_1-N_2)$ have $M\omega$ as the multiplication factor. From the above expressions, one can also calculate the real and imaginary parts of the ratio $\{N_1/(N_1-N_2)\}$:
\begin{equation}\label{Re_ratio}
\text{Re}\left(\frac{N_1}{N_1-N_2}\right)=\frac{(ma/M)^2\,x_{N_1}x_{N_{12}}+y_{N_1}y_{N_{12}}}{\{(ma/M)\,x_{N_{12}}\}^2+\{y_{N_{12}}\}^2}\,,
\end{equation}
and
\begin{equation}\label{Im_ratio}
\text{Im}\left(\frac{N_1}{N_1-N_2}\right) =\frac{ma}{M}\left[\frac{y_{N_1}x_{N_{12}}-x_{N_1}y_{N_{12}}}{\{(ma/M)\,x_{N_{12}}\}^2+\{y_{N_{12}}\}^2}\right]\,,
\end{equation}
which shows that the imaginary part of the ratio $\{N_1/(N_1-N_2)\}$ has $am$ as a multiplication factor. 

With the help of the above equations, we can further simplify the form of the TLNs, as presented in \ref{eq_39_a}. In particular, we can 
drop the second and the higher order terms of $M\omega$ in the square bracket of \ref{eq_39_a}, 
which leads to
\begin{align}
k_{lm}&=-P_{+}F_{2}\Bigg[-\pi\coth(2 P_+\pi)\text{Re}\left(\frac{N_1}{N_1-N_2}\right)\text{Re}N_{2}
+\text{Im}\left(\frac{N_1}{N_1-N_2}\right)\left\{1+\pi \coth(2 P_+\pi)\textrm{Im}N_{2}\right\}
\nonumber
\\
&+\text{Re}\left(\frac{N_1}{N_1-N_2}\right)\text{Im}(M\omega F_1)
+\text{Im}\left(\frac{N_1}{N_1-N_2}\right)\text{Re}(M\omega F_1)\Bigg]~.
\end{align}
Using the form of $P_+$, and the real and imaginary parts of $N_1$, $N_2$, and $\{N_1/(N_1-N_2)\}$ as derived above, we can rewrite the above expression as 
\begin{align}
k_{lm}&=-(am)^2\frac{F_{2}}{(r_+-r_-)M}\frac{y_{N_1}x_{N_{12}}-x_{N_1}y_{N_{12}}}{\{(ma/M)\,x_{N_{12}}\}^2+\{y_{N_{12}}\}^2}
\nonumber
\\
&-am\omega\frac{F_2}{r_+-r_-}\Bigg[\left\{-am\,x_{N_2}\pi\coth(2 P_+\pi)+\frac{\text{Im}(M\omega F_1)}{\omega}\right\}\frac{(ma/M)^2\,x_{N_1}x_{N_{12}}+y_{N_1}y_{N_{12}}}{\{(ma/M)\,x_{N_{12}}\}^2+\{y_{N_{12}}\}^2}
\nonumber
\\
&+\left\{am\,y_{N_2}\pi \coth(2 P_+\pi)+\frac{ma}{M}\frac{\text{Re}(M\omega F_1)}{\omega}-2r_+\right\}\,\frac{y_{N_1}x_{N_{12}}-x_{N_1}y_{N_{12}}}{\{(ma/M)\,x_{N_{12}}\}^2+\{y_{N_{12}}\}^2}\Bigg]\,,
\end{align}
where we dropped the additional second and higher order terms of $M\omega$.
Since $F_2$, $P_+$, and $M\omega F_1$ have $M\omega$ dependence in their expressions, 
we can further simplify the above expression and write it as Eq.~(7) of the main text, with $k^{(0)}_{lm}$ and $k^{(1)}_{lm}$ given by
\begin{equation}
k^{(0)}_{lm}=-\frac{\Tilde{F}_{2}}{(r_+-r_-)M}\frac{y_{N_1}x_{N_{12}}-x_{N_1}y_{N_{12}}}{\{(ma/M)\,x_{N_{12}}\}^2+\{y_{N_{12}}\}^2}
\end{equation}
and
\begin{align}
k^{(1)}_{lm}&=-\frac{\Tilde{F}_2}{r_+-r_-}\Bigg[\left\{-am\,x_{N_2}\pi\coth(2 P_+^A\pi)+\frac{\text{Im}(M\omega \Tilde{F}_1)}{\omega}\right\}\frac{(ma/M)^2\,x_{N_1}x_{N_{12}}+y_{N_1}y_{N_{12}}}{\{(ma/M)\,x_{N_{12}}\}^2+\{y_{N_{12}}\}^2}
\nonumber
\\
&+\Bigg\{am\,y_{N_2}\pi \coth(2 P_+^A\pi)+\frac{ma}{M}\frac{\text{Re}(M\omega \Tilde{F}_1)}{\omega}-2r_+
\\
\nonumber
&-\frac{4amr_+}{(r_+-r_-)}\sum_{k=1}^{l}\frac{4P_+^A}{k^2+4(P_+^A)^2}\Bigg\}\,\frac{y_{N_1}x_{N_{12}}-x_{N_1}y_{N_{12}}}{\{(ma/M)\,x_{N_{12}}\}^2+\{y_{N_{12}}\}^2}\Bigg]~,
\end{align}
where
\begin{equation}
P_+^A = \frac{am}{r_+-r_-}, \qquad \Tilde{F}_2 = \frac{ \Gamma \left(-1+l\right)\Gamma \left(3+l\right)}{\Gamma \left(2 l+2\right) \Gamma \left(2 l+1\right) }\prod_{j=1}^{l} \left[j^2+4(P_+^A)^2\right]\,,
\end{equation}
and 
\begin{multline}
M\omega \Tilde{F}_1=N_1\psi(-1 +l)+N_1\psi(1 + l+2 i P_+^A)- N_2\psi(3 + l)-N_2\psi(1 + l-2 i P_+^A)
\\
-(N_1 - N_2)\psi(2 + 2 l)- (N_1 - N_2)\psi(1 + 2 l)\,.
\end{multline}
Note that $P^A_+$ and 
$\Tilde{F}_2$ are independent of $\omega$. Therefore, 
$k^{(0)}_{lm}$ and $k^{(1)}_{lm}$ are independent of $\omega$ as well.
Moreover, $k^{(0)}_{lm}$ and $k^{(1)}_{lm}$ remain unchanged under the transformation $(a, \omega)\to (-a, -\omega)$, which implies  that $k_{lm}$ will also remain the same under that transformation. Thus, the leading terms in $k_{lm}$ of Eq. (7) in the main text are consistent with the expectation that TLNs are associated with conservative terms of the tidal response.

\section{Tidal response of an extremal Kerr black hole}\label{supp_mat:sec_2}

In this section, we will present the tidal response of an extremal Kerr BH and shall analyze its real part, which determines the dynamical TLNs as noted above. Since the coordinate $z=(r-r_+)/(r_+-r_-)$, defined in the previous section, is not well behaved for an extremal Kerr BH, with $a=M$ (i.e., $r_{+}=r_{-}$), we introduce a new coordinate $\bar{z}\equiv (r-r_+)/r_+$. This coordinate transformation simply shifts the origin of the radial coordinate to the position of the event horizon. 
Thus, the source-free radial Teukolsky equation in the new coordinate $\bar{z}$ becomes
\begin{equation}\label{eq_extreme_1}
\frac{\mathrm{d}^2R}{\mathrm{d}\bar{z}^2}+\left[-\frac{2(1+2iM\omega)}{\bar{z}}+\frac{2i(m-2M\omega)}{\bar{z}^2} - 2iM\omega\right]\frac{\mathrm{d}R}{\mathrm{d}\bar{z}} 
+\left[\frac{-2iM\omega}{\bar{z}}+\frac{-\lambda+6iM\omega}{\bar{z}^2}-\frac{8 i (m-2 M \omega )}{\bar{z}^3}\right]R=0~.
\end{equation}
In the near-zone regime ($M\omega\bar{z}\ll 1$) and using the small frequency approximation ($M\omega\ll1$), the Teukolsky equation, as presented in \ref{eq_extreme_1}, can be re-written as
\begin{equation}\label{master_equation_ext_BH}
\frac{\mathrm{d}^2R}{\mathrm{d}\bar{z}^2} + \left[\frac{2(-1-2iM\omega)}{\bar{z}} + \frac{2i(m-2M\omega)}{\bar{z}^2}\right]\frac{\mathrm{d}R}{\mathrm{d}\bar{z}} \\+\left[\frac{-\lambda +6 i M \omega}{\bar{z}^2}-\frac{8 i (m-2 M \omega )}{\bar{z}^3}\right]R = 0~,
\end{equation}
where $\lambda$ is the separation constant between the radial and the angular equations, which reads
\begin{equation}\label{def_lambda_1}
\lambda=l(l+1)-2-2mM\omega\left\{\frac{4}{l(l+1)}+1\right\}\,.
\end{equation}
Here, we have neglected all the second and higher order terms of $M\omega$. Note that in order to arrive at \ref{master_equation_ext_BH}, which is our master equation for extremal Kerr BH, we have ignored the $-2iM\omega$ term in the coefficient of $R'(\bar{z})$ and $(-2iM\omega/\bar{z})$ in the coefficient of $R(\bar{z})$, since both of them contribute at $\mathcal{O}(\bar{z}^{2})$. 

Since the master equation
in \ref{master_equation_ext_BH} has a regular singular point at $\bar{z}=\infty$ and an irregular singular point of rank one at $\bar{z}=0$, its solution can be written in terms of the confluent hypergeometric function:
\begin{multline}
R(\bar{z})=\bar{z}^{\frac{3+4iM\omega-\beta}{2}}\left[c_{1}U\left(\frac{5-4iM\omega+\beta}{2},1+\beta,\frac{2i(m-2M\omega)}{\bar{z}}\right) \right.
\\
\left.+ c_2 e^{\frac{2i(m-2M\omega)}{\bar{z}}} U\left(\frac{-3+4iM\omega+\beta}{2},1+\beta,-\frac{2i(m-2M\omega)}{\bar{z}}\right)\right]~,
\end{multline}
where $c_1$ and $c_2$ are the constants of integration, and the quantity $\beta$ is defined as
\begin{align}\label{def_beta_a}
\beta&\equiv
2l+1-2N+\mathcal{O}(M^{2}\omega^{2})~;
\quad
N\equiv \frac{2mM\omega\left(l^2+l+4\right)}{l(l+1)(2l+1)}~.
\end{align}
In order to determine the tidal response function,
we must fix one of the arbitrary constants $c_{1}$ and $c_{2}$ by imposing appropriate boundary condition. Since we are dealing with BHs, as before, the natural boundary condition corresponds to the absence of outgoing waves at the horizon. It demands $c_2=0$ and, hence, the radial Teukolsky function becomes
\begin{equation}\label{solution_final_ext_BH}
R(\bar{z})=c_1 \bar{z}^{\frac{3+4iM\omega-\beta}{2}} U\left(\frac{5-4iM\omega+\beta}{2},1+ \beta,\frac{\mathcal{C}}{\bar{z}}\right)~.
\end{equation}
To calculate the tidal response function, we need the large $r$ limit (or, large $\bar{z}$ limit) of the radial Teukolsky function. From which, we can read off the tidal response function as the coefficient of $\bar{z}^{-\beta}$. Note that here also, to leading order in $M\omega$, the fall-off of the radial Teukolsky function corresponds to $\bar{z}^{-2l-1}(1+2N\ln \bar{z})$, and following the 
previous section,
we leave out the $\ln \bar{z}$ term from the expression for the tidal response function. 
Therefore, the dynamical tidal response function of an extremal Kerr BH under external perturbation, can be written as
given in Eq.~(8) of the main text.

Similar to the case of the non-extremal Kerr BH, even in the extremal limit the dynamical  (i.e., $M\omega\ne0$) 
case must be dealt with separately. Using the fact that $N$ is a linear function of $M\omega$ and $l\in\mathbb{Z}_{\ge2}$, the dynamical response function (Eq.~(8) of the main text) can be simplified to the following expression 
when retaining terms upto linear order in $M\omega$: 
\begin{multline}\label{simplified_non_static_rf_ext_BH}
F_{lm}=(-1)^{l}i^{2l+1-2N}\left\{2(m-2M\omega)\right\}^{2l+1-2N}\left(\frac{\sin[(2iM\omega-N)\pi]}{\sin[(2N)\pi]}\right)\frac{\Gamma\left(-1+l\right) \Gamma \left(3+l\right)}{\Gamma \left(1+2l\right)\Gamma \left(2l+2\right)}
\\
\times\left[1+(2 i M \omega-N)\psi(-1+l)+(-2 i M \omega-N)\psi(3+l)+2N\psi(1+2l)+2N\psi(2+2l)\right]\,.
\end{multline}
In the small frequency limit ($M \omega \ll 1$), 
the parenthetic ratio of the sines in the above expression can be expanded as
\begin{equation}\label{def_alpha_1b}
\frac{\sin[(2iM\omega-N)\pi]}{\sin[(2N)\pi]}\simeq \frac{2iM\omega-N}{2N}
=\frac{1}{2}\left[-1+i\frac{l(l+1)(2l+1)}{m\left(l^2+l+4\right)}\right]\,.
\end{equation}
Note that $M\omega$ is non-zero in the above expression.
We can further simplify $i^{2l+1-2N}=(-1)^{l}ie^{-iN\pi}\approx(-1)^{l}i\left[1-iN\pi\right]$
since $M\omega$ is assumed to be small in our calculations. 
Therefore, the dynamical response function for an extremal Kerr BH under an external perturbation becomes
\begin{multline}\label{simplified_non_static_rf_ext_BH_02}
F_{lm}=-i\left(1-iN\pi\right)(2m)^{2l-2N}\left\{m-(2l+1)2M\omega\right\}\left[1-i\frac{l(l+1)(2l+1)}{m\left(l^2+l+4\right)}\right]\frac{\Gamma\left(-1+l\right) \Gamma \left(3+l\right)}{\Gamma \left(1+2l\right)\Gamma \left(2l+2\right)}
\\
\times\Big[1+(2 i M \omega-N)\psi(-1+l)+(-2 i M \omega-N)\psi(3+l)+2N\psi(1+2l)+2N\psi(2+2l)\Big]\,.
\end{multline}
Given the fact that $N$ is real, one can read off the real part of the above dynamical tidal response function and,  hence, obtain the dynamical TLNs to be
\begin{align}\label{tlnextremalbh}
k_{lm}&=(2m)^{2l-2N-1}\left\{m-(2l+1)2M\omega\right\}\left(\frac{\Gamma\left(-1+l\right) \Gamma \left(3+l\right)}{\Gamma \left(1+2l\right)\Gamma \left(2l+2\right)}\right)
\nonumber
\\
&\times \Bigg[-\frac{l(l+1)(2l+1)}{\left(l^2+l+4\right)}\left\{1+N\left(2\psi(1+2l)+2\psi(2+2l)-\psi(-1+l)-\psi(3+l)\right) \right\}
\nonumber
\\
&\qquad \qquad +2Mm\omega \left\{\psi(-1+l)-\psi(3+l)-\frac{\pi m \left(l^2+l+4\right)}{l(l+1)(2l+1)}\right\}\Bigg]\,.
\end{align}
Since the above expression contains some second and higher order terms of $M\omega$, we can further simplify it and write it as the Eq.~(10) of the main text, with $\Tilde{k}^{(0)}_{lm}$ and $\Tilde{k}^{(1)}_{lm}$ given as
\begin{equation}
    \Tilde{k}^{(0)}_{lm} = -\frac{2l(l+1)(2l+1)}{\left(l^2+l+4\right)}\left(\frac{\Gamma\left(-1+l\right) \Gamma \left(3+l\right)}{\Gamma \left(1+2l\right)\Gamma \left(2l+2\right)}\right),
\end{equation}
and
\begin{align}
\Tilde{k}^{(1)}_{lm}&=\frac{\Gamma\left(-1+l\right) \Gamma \left(3+l\right)}{\Gamma \left(1+2l\right)\Gamma \left(2l+2\right)}\Bigg[-\frac{2lm(l+1)(2l+1)}{\left(l^2+l+4\right)}\left\{\frac{N}{M\omega}\left(2\psi(1+2l)+2\psi(2+2l)-\psi(-1+l)-\psi(3+l)\right) \right\}
\nonumber
\\
& +4m^2 \left\{\psi(-1+l)-\psi(3+l)-\frac{\pi m \left(l^2+l+4\right)}{l(l+1)(2l+1)}\right\}
+
\frac{4l(l+1)(2l+1)^2}{\left(l^2+l+4\right)}\Bigg]
\,.
\end{align}

\subsection{Tidal response of an extremal Kerr black hole in an axi-symmetric tidal field}\label{app_B.1}

In this section, we will present an alternative expression for the tidal response function of an extremal Kerr BH in an axi-symmetric tidal background ($m=0$) . For axi-symmetric tidal perturbation, the dynamical tidal response function (Eq.~(8) of the main text)
takes the following form,
\begin{equation}\label{resp_func_ext_BH_axi_sym_11}
F_{lm}=-(-4iM\omega)^{2l+1}\frac{\Gamma \left(-2l\right) \Gamma \left(3+l-2 i M \omega\right)}{\Gamma \left(2-l-2 i M \omega\right)\Gamma \left(2l+2\right)}\,.
\end{equation}
Expanding all relevant Gamma functions in the expression above in  powers of $M\omega$, we obtain
\begin{equation}\label{simplify_ext_BH_axi_sym_1}
F_{lm}=-(-4iM\omega)^{2l+1}\frac{\Gamma \left(-2l\right) \Gamma \left(3+l\right)}{\Gamma \left(2-l\right)\Gamma \left(2l+2\right)}\left[1-2iM\omega\,\psi(3+l)+2iM\omega\,\psi(2-l)\right]\,,
\end{equation}
where we have neglected the second and higher order terms in $M\omega$ inside the square brackets. Since our calculation is valid upto the linear order of $M\omega$, we can neglect the $\mathcal{O}(M\omega)$ terms inside the square bracket in the above expression.
Thus, for $l\in \mathbb{Z}_{\ge2}$, \ref{simplify_ext_BH_axi_sym_1} becomes
\begin{equation}
F_{lm}=-\frac{(-1)^l}{2}(-4iM\omega)^{2l+1}\frac{\Gamma\left(-1+l\right) \Gamma\left(3+l\right)}{\Gamma\left(2l+1\right)\Gamma\left(2l+2\right)}\,.
\end{equation}
We notice that there is an overall
multiplicative factor of $\sim(-4iM\omega)^{2l+1}$ in the above equation. Also, the dynamical response function is purely imaginary in nature, signaling vanishing TLNs for extremal Kerr BHs under axi-symmetric tidal perturbations. Moreover, since the tidal response function has $(M\omega)^{2l+1}$ as a multiplicative factor, which is $(M\omega)^{5}$ for $l=2$, it follows that at linear order in $M\omega$, the tidal response function identically vanishes. 

There is another alternative derivation of this result in which one assumes $l\in\mathbb{Z}_{\ge2}$ from the beginning and writes down the solution of the Teukolsky equation for $m=0$ as
\begin{equation}
R(z) = c_1 \bar{z}^{-l+1+2iM\omega} U\left(3+l-2iM\omega,2l+2, -\frac{4iM\omega}{\bar{z}}\right)\,.
\end{equation}
If we take $l\in\mathbb{Z}_{\ge2}$, then $2l+2$ is an integer and greater than or equal to $4$. Thus the large $r$ limit of the above solution to the radial Teukolsky equation is, 
\begin{equation}
R(z) \sim c_1 \bar{z}^{l+2+2iM\omega} \frac{\Gamma(2l+1)}{\Gamma(-1+l-2iM\omega)}(-4iM\omega)^{-2l-1}\left[1+ 0\times \bar{z}^{-2l-1}\right]\,,
\end{equation}
which too implies a vanishing tidal response function. All of these results are consistent with the one presented in the main text. 


\putbib[reference_arXiv_SM]

\end{bibunit}
\end{document}